\documentclass{article}
\usepackage{fleqn,ams}
\usepackage{amssymb}
\usepackage{epsfig}
\usepackage{verbatim}
\usepackage[cp1251]{inputenc}
\newcommand{\const}{\mbox {Const}}
\newcommand{\ris}[2]{\refstepcounter{figure}\begin{center}
\begin{tabular}{c}
\parbox{10cm}{\epsfig{file=#1,width=8cm}}\\[12pt]
\parbox{10cm}{\noindent Figure: \thefigure. {\sl #2}}\\
\end{tabular}
\end{center}}

\begin{document}

\begin{center}
{\bf\Large Gravimagnetic shock waves
and gravitational-wave experiments} \\[12pt]
Yu.G. Ignat’ev \\
Kazan State Pedagogical University\\ 1 Mezhlauk Str., Kazan
420021, Russia
\end{center}

\begin{abstract}
Causes of the unsatisfactory condition of the gravitational-wave
experi\-ments are discussed and a new outlook at the detection of
gravitational waves of astrophysical origin is proposed. It is
shown that there are strong grounds for identifying the so-called
giant pulses in the pulsar NP 0532 radiation with gravimagnetic
shock waves (GMSW) excited in the neutron star magnetosphere by
sporadic gravitational radiation of this pulsar.
\end{abstract}

\section{Introduction}
In the history of physics of the 20th century, I suppose, there is
no such a grave experimental problem (except the controlled
thermonuclear fusion problem) that, being solved for over thirty
years by different research groups as the gravitational wave (GW)
detection problem. Although much means are used to solve it, no
sufficiently convincing possitive results have been obtained. What
are the reasons for this situation? An error in the gravitational
theory? The experimentalists’ incapability? Are there realizable
opportunities to detect gravi\-ta\-ti\-o\-nal radiation in the
visible future? Is the GW detection problem worth studing? We will
try to answer these questions in the present paper. As any other
radiation detection problem, this one also splits into two
independent problems: (1) “GW sources” and (2) “GW detectors”. It
is important not to forget to join both branches by solving a
concrete experimental problem.

\section{Sources of gravitational radiation}
\subsection{Estimates of gravitational radiation power}
The average power of gravitational radiation from a source is
calculated by the formula  \cite{torn}
\begin{equation}  \label{GSMW555}
L_{GW} = \frac{G}{5c^5}< \stackrel{...}{t} _{ik}\stackrel{...}{t}
_{ik}>,
\end{equation}
where $G$ is the gravitational constant, $c$ is the velocity of
light,
\begin{equation}  \label{GSMW556}
t_{ik} = \int \rho(x_ix_k - \frac{1}{3}\delta_{ik}r^2)dV
\end{equation}
is the reduced quadrupole moment of the source; dots mean time
derivatives. There are two parameters of interest in the GW
detection problems: the GW magnitude (i.e. the deviation from the
flat metric $h_{ik} = g_{ik} - \eta_{ik}$) and the GW frequency,
$\omega$. The GW energy flow is expressed in terms of these
parameters by the formula \cite{torn}:
\begin{equation}  \label{GSMW557}
ct^{14} = {\cal P} = \frac{c^3}{16\pi G}\left[\dot{h}^2_{23} +
\frac{1}{4}(\dot{h}_{22} - \dot{h}_{33})^2\right].
\end{equation}
(Throughout the paper we use the metric signature $(-1,-1,-1,
+1)$). From (1) - (3) follow estimating formulae for the
gravitational radiation power and magnitude:
\begin{equation}  \label{GSMW558}
L_{GW} = \frac{GE^2_q\omega^2}{5c^5};
\end{equation}
\begin{equation}  \label{GSMW559}
L_{GW} = \frac{c^3\omega^2R^2}{4G}h^2,
\end{equation}
where $E_q$ is the energy of quadrupole ole oscillations of the
source of the characteristic frequency $\omega$ ($L_0 = \omega
E_q$ is the quadrupole oscillation power), R is the distance from
the source to an observer.

\subsection{Restrictions on GW magnitude}
In particular, a useful formula follows from (\ref{GSMW558}) --
(\ref{GSMW559}):
\begin{equation}  \label{GSMW560}
\frac{h}{h_0} = \frac{E_q}{Nc^2},
\end{equation}
where $h_0$ is the gravitational potential of the source of the
total mass $M$:
\begin{equation}  \label{GSMW561}
h_0 = \frac{GM}{c^2R}.
\end{equation}
According to (\ref {GSMW560}) and (\ref{GSMW561})), the ratio ofGW
magnitude to the Newtonian gravi\-ta\-tio\-nal potential of the
source is of the order of the ratio of the quadrupole energy of
the source oscillations to its complete energy at rest, $E_0 =
Mc^2$. It is obvious that always $E_q <E_0$, and $E_q/E_0 \ll 1$
in a generic situations, therefore Eq. (\ref{GSMW561}) gives an
upper limit of the GW magnitude of a source, which is a good
sobering factor by itself. Let us list some values for reference.
For the Solar mass ($M_{\odot} = 2\cdot 10^{33}$g) and a distance
of 1 pc (3.26 light years = $3\cdot 10^{18}$ cm)1, from Eq.
(\ref{GSMW561}) we obtain\footnote{The nearest stars’ distance is
about 1,3 pc.}:
\begin{displaymath}
h_0\cdot(\mbox{pc}/M_{\odot}) = 4,8\cdot 10^{-14}.
\end{displaymath}
For a mass of 1 kg at a distance of 1 m:
\begin{displaymath}
h_0\cdot (\mbox{m kg}^{-1}) = 7,4\cdot 10^{-18}.
\end{displaymath}
Therefore for an A-bomb explosion with $(\Delta M/M\sim 10^{-3})$,
under the condition that the whole explosion energy turns into the
quadrupole oscillations energy, at a distance of $1$ m (!) from
the epicentre, we get the GW magnitude $h \sim 10^{-21}$. If one
makes all atoms of a compact graser oscillate in the optic range
(the radiation energy is about ($\hbar \nu \sim 1$ ev, $l\sim 1$
m), we get the following maximum estimate for the GW magnitude on
the graser end-wall: $h\sim 10^{-28}$.

GW sources can be divided into two classes: (1) stable
(quasistable) sources, which cannot be destroyed during the GW
radiation process; (2) catastrophic sources, being destroyed in
the GW radiation process. A graser represents a source of the
first type, an A-bomb a source of the second type. For first type
sources the quadrupole oscillation energy cannot exceed the
binding energy of the source as a whole, unlike second-type ones,
which are sources for one occasion. For example, close binaries
and quadrupole oscillations of neutron stars are first-type
astrophysical sources, and Supernovae are second-type ones.

As mentioned above, stable radiation sources are subject to the
condition
\begin{equation}  \label{GSMW562}
E_{\mbox{kin}} < E_b,
\end{equation}
where $E_{\mbox{kin}}$ is the inner kinetic energy of separate
parts of the source, $E_b$ is their binding energy. Since it is
always $E_q \leq E_{\mbox{kin}}$, the condition (\ref{GSMW562}
takes the form
\begin{equation}  \label{GSMW563}
E_q \leq E_b.
\end{equation}
Hence the upper limit of GW magnitude from such sources can be
obtained from the formula
\begin{equation}  \label{GSMW564}
h < h_0\frac{E_b}{Mc^2}.
\end{equation}
For astrophysical sources the binding energy is essentially that
of gravitational attraction. Let $\Delta M$ be the part of the
mass of an astrophysical object performing quadrupole
oscillations. Its gravitational binding energy is
\begin{equation}  \label{GSMW565}
E_b < G\frac{\Delta M\cdot M}{l},
\end{equation}
where $l$ is the chracteristic size of the system. Thus for the
upper limit of a GW magnitude from such a source Eqs.
(\ref{GSMW564}) and (\ref{GSMW565}) give:
\begin{equation}  \label{GSMW566}
h\leq h_0\frac{r_g}{2l}\frac{\Delta M}{M},
\end{equation}
where $r_g = 2GM/c^2$ is the gravitational radius of the radiating
system.

\subsection{Radiation frequency}
Consider first a source of total mass $M$, which consists of two
parts, so that the second part $\Delta M$ performs a free motion
in the gravitational field of the system (rotation or free fall).
Let $\omega$ be a characteristic frequency of this
process\footnote{Evidently the order of magnitude of this quantity
coincides with the frequency of gravitational radiation from the
system.}. Equating the centrifugal and free fall accelerations, we
obtain the well-known relation
\begin{equation}  \label{GSMW567}
GM = \omega^2l^3,
\end{equation}
which connects the characteristic size of the system with its
characteristic fre\-qu\-en\-cy.

Now let the gravitational attraction in the system be held by the
forces of pressure (for stellar quadrupole oscillations). Equating
these forces, we get the hydrostatistic balance condition
\begin{equation}  \label{GSMW568}
|\vec{\nabla}P| = \rho \frac{G M}{l^2},
\end{equation}
where  $P$ is the pressure and $\rho$ is the density. Using the
known relation$dP = v^2_fd\rho$, where $v_f$ is the velocity of
sound, from Eq. (\ref{GSMW568}) we obtain:
\begin{equation}  \label{GSMW569}
lv^2_f\approx MG.
\end{equation}
But $v_f/l\approx \omega$ is the system proper oscillation
frequency. Therefore for systems supported by the forces of
pressure we return to the estimate of Eq. (13).

{\it  Thus for stable astrophysical GW sources Eq. (13) has a
universal nature if $omega$ is understood as a characteristic
frequency of the system oscillations}.

From the law (13) we can estimate the radiation characteristics of
the col\-lap\-sing objects, colliding stars and the like. It
follows from this law that a maximum radiation frequency can be
achieved for objects close to the gravitational collapse
condition. In this case, by (12), the maximum magnitude of
radiated GW is achieved (see \cite{1}). For objects of masses of
the order of the Solar mass ($r_g =2,96$ km) the maximum radiation
frequency is
\[\omega_{max}\sim c/r_g\approx 10^5\mbox{sec}^{-1}.\]

\subsection{Close binary stellar systems}
For intense astrophysical sources of gravitational radiation, this
radiation is the basic mechanism of quadrupole oscillation energy
loss. Therefore, more rigorously, such sources should be called
quasistable. The gravitational radiation power of a system of two
orbiting gravitating masses m1 and m2 is calculated from the known
formula \cite{land}
\begin{equation}  \label{GSMW570}
L_g = -\frac{dE}{dt} = \frac{32G^4m^2_1m^2_2(m_1 + m_2)}{5c^5r^5},
\end{equation}
where $r$ is the separation of the centres of mass. The energy
balance leads to the mass approaching law \cite{land}
\begin{equation}  \label{GSMW571}
\dot{r} = \frac{64G^3m_1m_2(m_1 + m_2)}{5c^5r^3},
\end{equation}
Its integration yields a formula for the time $t$ needed for the
mass centres to approach to a distance of $r$ from $r_0$:
\begin{equation}  \label{GSMW572}
t = \frac{5c^5}{192G^3m_1m_2(m_1 + m_2)}(r^4 - r_0^4).
\end{equation}
Further for simplicity we will study a pair of equal stars,
setting $m_1 = m_2 = M$, $r_0 = 2R_0$ where $R_0$ is the stellar
radius, i.e. we will calculate the time until the catastrophic
stellar collision, $\tau$ (the lifetime). Then from
(\ref{GSMW572}) -- (\ref{GSMW572}) we get:
\begin{displaymath}
\tau = \frac{5c^5(l^4 - 16R^4_0)}{384G^3M^3}
\end{displaymath}
and for $l\gg R_0$
\begin{equation}  \label{GSMW573}
\tau \approx \frac{5}{384}\left(\frac{l}{r_g}\right)^3\frac{l}{c}.
\end{equation}
The gravitational radiation frequency of a binary system increases
with time; the ratio of the frequency shift per period $\Delta
\omega$ to the radiation frequency $\omega$ is, by order of
magnitude,
\begin{equation}  \label{GSMW574}
\frac{\Delta \omega}{\omega} \sim
\left(\frac{r_g}{l}\right)^{\frac{5}{2}}.
\end{equation}

Figs. 1 and 2 show the dependence of the gravitational radiation
power of a binary system and the radiated GW magnitude on the
distance between the stars. Fig. 3 shows the binary lifetime
versus their separation for stars with the masses $m_1 = m_2 =
M_\odot$.

\ris{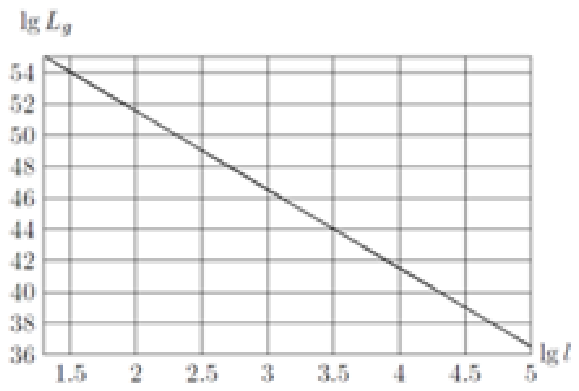}{Gravitational radiation power of a binary $L_g$
(erg/s) vs. distance between the stars $l$ (km).}

\ris{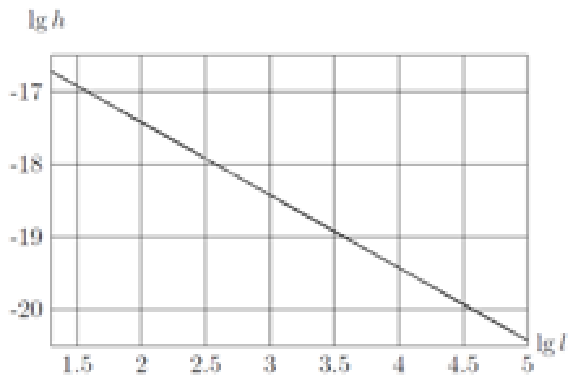}{GW magnitude h from a binary vs. its size $l$
(km), at a distance of 1 kpc}

Let us estimate the probability of {\it GW detection from a close
binary system} in the Galaxy at given rotation period, assuming
that the Galaxy age is of the order of $1\cdot 10^{10}$ years.
Further, we assume that the average stellar number density in the
Galaxy is of the order of $0,120$ stars/pc$^3$ \cite{15}, the
Galactic volume is 300 kpc$^3$ \cite{16}, then the number of stars
in the Galaxy is about $0,35 \cdot 10^{11}$. Besides, we take into
account that approximately half of the stars are in binary systems
\cite{17}. Then the possibility of existence of a binary with a
prescribed lifetime $\tau$ is proportional to the ratio $\tau/t$,
where $t$ is the age of the Galaxy. In the columns of Table 1
corresponding to system lifetimes smaller than 1 year, the
detection probability of such systems in experiments lasting 1
year is shown. Evidently for such systems the probability of
detection in a year-lasting experiment coincides with that for a
binary having a lifetime of 1 year. The number of such systems in
the Galaxy is estimated to be of the order of one.

\ris{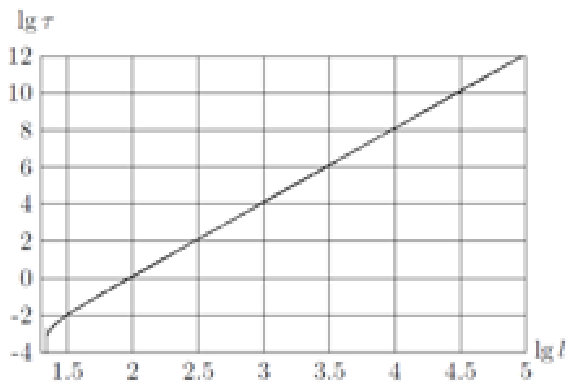}{Lifetime of a binary $\tau$ (seconds) vs. its size
$l$ (km)}

The presented data show that at the instant preceding the
catastrophic collision, the gravi\-ta\-tio\-nal radiation power
from the binary is of the order of Supernova luminosity. Thus, a
stellar collision in a close binary is an event whose scale is of
the order of a Supernova explosion. As mentioned above, in the
Galaxy the probability of GW detection from a binary with a
lifetime of the order of1 year is close to one. This means that
catastrophic phenomena with energy release of the order of $1\cdot
10^{54}$ erg/s should happen once a year. However, in reality such
phenomena happen once in 40 to 80 years in the Galaxy \cite{19}. A
reason for such a discrepancy is that in a close binary with a
lifetime of the order of1 year the stars’ separation is about
12000 km.

Therefore for such a system to exist it is necessary that both
stars be at least white dwarfs. But in this case even in a stellar
collision the energy released is 4 orders of magnituded smaller
than that of a Supernova explosion. Thus for a catastrophic
collision of this scale it is at least necessary that one of the
components be a neutron star, while the second one is a white
dwarf. The existence probability of such systems in the Galaxy is
much smaller. Note that it is difficult to understand the
experimental programmes intended for registration of GW from the
binaries with periods of the order of a few seconds. According to
Table 1, their lifetime does not exceed 5 years, and in this case
it would be more reasonalble to wait these 5 years and to detect
the gravitational radiation from a catastrophic collision: its
power is higher by at least 13 orders and the GW magnitude is
greater by 3 orders (!), as follows from Table 1. However, at
least in the last 10 years nobody detected catastrophic events on
such a scale at distances smaller than 15 kpc.

Since the pulsars are identified with Supernovae remnants, the
average frequency of Supernova bursts may be estimated from the
data on pulsars spreading in the Solar neighbourhood. Thus, at
distances within about 1 kpc, on the whole, about 20 pulsars are
observed. Table 2 shows the data on the pulsars nearest to the
Solar system3. As follows from this Table, almost all the pulsars
are younger than 108 years. Therefore, it may be stated that the
observed pulsars are remnants of Supernovae which exploded in the
last hundred million years. It gives 1 flash per 50 years,
coinciding with the estimate of Ref . \cite{torn}. It seems likely
to be also close to the average frequency of catastrophic
collisions in close binaries.\pagebreak

%
%\vskip 12pt \hrule \vskip 12pt

\refstepcounter{table} Table \thetable: {\sl Characteristics of
gravitational radiation from a close binary$^*$}
\begin{flushleft}
{\footnotesize
\begin{tabular}{|l|l|l|l|l|l|l|l|l|l|l|}
\hline
$l$ & 5(4)& 2(4)& 1(4)& 7(3)& 4(3)& 1(3)& 320 & 100 & 40 & 20\\
\hline
$\tau$ & 8,2(10)& 2,1(9)& 1,3(8)& 3,15(7)& 3,3(6) & 1,3(4) & 129 &
1,30 & 0,031 & 0\\
\hline
$\omega$ & 0,046& 0,18& 0,52& 0,88 & 2,05& 16,4 & 92,6& 518& 2047&
5790\\
\hline
$T$ & 136& 34,3& 12,1& 7,10& 3,07& 0,38& 0,068& 0,012& 0,003&
0,001\\
\hline
$L_g$ & 1,1(38)& 1,1(40)& 3,4(41)& 2,0(42)& 3,3(43)& 3,4(46) &
1,1(49)& 3,4(51)& 3,3(53)& 1,1(55)\\
\hline
$h$ & 7,5(-21)& 1,9(-20)& 3,7(-20)& 5,4(-20)& 9,4(-20)& 3,7(-19)&
1,2(-18) & 3,7(-18)& 9,4(-18)& 1,9(-17)\\
\hline
${\cal P}$ &9,7(-14)& 9,5(-12) & 3,0(-10)& 1,8(-9)& 3,0(-8)
&3,0(-5)&
9,8(-3) & 3,0 & 296 & 9463\\
\hline
$N$ & 2343 & 60 & 3,7 & 1 & 1 & 1 & 1 & 1 & 1 & 1\\
\hline
$h_\oplus$& 1,5(-20)& 4,4(-21)& 1,7(-21)& 3,7(-21)& 6,5(-21)&
2,6(-20)& 8,3(-20)& 2,6(-19)& 6,5(-19)& 1,3(-18)\\
\hline
${\cal P}_\oplus$ &  3,8(-13) & 3,2(-12)& 1,6(-11) & 9,0(-12)&
1,5(-10)&
1,5(-7) &4,7(-5)& 1,5(-2) & 1,4 & 46\\
\hline
\end{tabular}}\\[12pt]
{\small\sl  $^{*)}$ Here and henceforth, the figures in
parantheses indicate the order of magnitude (a(b) = a$\cdot
10^b$); the quantities $h$ and ${\cal P}$ (the gravitational
radiation flow density) are calculated at a distance of 1 kpc from
the binary; $T=2\pi/\omega$ is the GW period; N is the expected
number of binaries in the Galaxy; $R = <R>$ is the expected
distance to the binary in kpc; $h_\oplus$ is the expected GW
magnitude on the Earth; P. is the expected gravitational radiation
flow density on the Earth. The quantities $T$  and $\tau$ are
given in seconds, $\omega$  in sec$^{-1}$, $l$ in km, $L_g$ in
erg/sec, ${\cal P}$ and ${P}_\oplus$ in W/cm$^3$.}\\[2pt]\hrule
\end{flushleft}

\vskip 12pt \refstepcounter{table} Table \thetable: {\sl Data on
pulsars located at distances smaller or of the order of 1 kpc from
the Sun}
\begin{center}
{\footnotesize
\begin{tabular}{|l|l|l|l|l|}
\hline
 & & & & \\[-4pt]
\parbox{0.75cm}{No\\} & \parbox{0.75cm}{Pulsar\\(name)} & \parbox{1cm}{Distance \\(kpc)} & \parbox{0.75cm}{P\\
(sec)} &
\parbox{1cm}{$\dot{P}/P$
(years}\\[8pt]
\hline
 & & & & \\[-4pt]
1 & MP 0031 & 0,21 & 0,94 & 7,1(7) \\[4pt]
2 & MP 0450 & 0,33 & 0,55  & \\[4pt]
3 & NP 0532 & 2,0 & 0,033 & 2,5(3)\\[4pt]
4 & MP 0628-28 & 0,170 & 1,24 & 1,6(7) \\[4pt]
5 & CP 0809 & 0,19  & 1,29 & 2,5(8) \\[4pt]
6 &AP 0823+26 0,38& 0,53& 1,0(7)\\[4pt]
7 & PSR  0833-45 & 0,5 &0,089 &2,3(4)\\[4pt]
8 &CP 0834 &0,43 &1,27 &5,9(6) \\[4pt]
9 &PP 0943& 0,30 &1,098& \\[4pt]
10& CP 0950 &0,10& 0,253& 3,5(7) \\[4pt]
11 &CP 1133 &0,16 &1,188& 1,0(7) \\[4pt]
12 &AP 1237+25& 0,20 &1,382& 4,6(7) \\[4pt]
13 &PSR 1451-68 &0,40 &0,263 >2,8(6)& \\[4pt]
14 &HP 1508 &0,26& 0,740 &4,7(6) \\[4pt]
15& CP 1919 &0,42& 1,337 &3,2(7) \\[4pt]
16 &PSR 1929+10& 0,27 &0,227& 6,2(6) \\[4pt]
17 &JP 1933+16 &3& 0,359 &1,9(6) \\[4pt]
18 &AP 2016+28& 0,47 &0,558& 1,2(8) \\[4pt]
19 &PSR 2045& 0,38&1,962 &5,7(6)\\[4pt]
\hline
\end{tabular}}\\[12pt]
\end{center}

Thus we have to deal at best with system linear sizes of the order
of 1$10000\div 20000$km. It gives: $T\sim 10\div 40$sec,
$h_{\oplus}\sim (1\div 5)\cdot 10^{-21}$, ${\cal P}_{\oplus}\sim
10^{-11}\div 3\cdot 10^{-12}$W/cm$^2$. GW with such parameters can
hardly be detected in the coming decades. In this situation it
only remains to hope for a case, rare and simultineously dangerous
for the Earth, of a Supernova burst or a catastrophic end of a
close binary.

\subsection{Neutron star oscillations}
There is, however, one more class of stable astrophysical GW
sources — quadrupole oscillations of neutron stars. Table 3 shows
the calculated parameters of gravitational radiation from neutron
stars \cite{torn} (columns 1 $\div$ 7). According to this table,
the following characteristics of radiation are to be expected from
these sources: $\sim (0,3\div 1)\cdot 10^4$s$^{-1}$; $h_\oplus
\sim 10^{-25}$, ${\cal P}_\oplus \sim 3\cdot 10^{-12}$W/cm$^2$ by
the energy ofstellar quadrupole oscillations of the order of
$10^{38}\div10^{39}$ erg and the whole gravitational luminosity of
$L_g \sim 4\cdot106{39}$ erg/sec. Note that due to the smallness
of the GW magnitude from this source and the absence of a
mechanism able to support the excitation of a necessary quadrupole
moment during a sufficiently long time, quadrupole oscillations of
neutron stars have not been considered as a competitive GW source.

\section{GMSW and GW detection}
The cause of the unsatisfactory condition in the GW detection
problem is, in the author’s opinion, the originally chosen
erroneous way ofits solution — the programme ofcreating GW
detectors. Direct GW detection can be realized either due to their
tidal effect on a nonrelativistic (solid-state) detector, or due
to their relativistic effect on a detector having a relativistic
component (a laser ray). In both cases the GW effect on a detector
(test-body displacement or laser ray deviation) is propotional to
the GW magnitude. And the expected GW magnitudes from
astrophysical sources are extremely small (see, for example,
\cite{torn}).

The existing GW detection programmes are generally meant for
astrophysics sources of two types: $1.$ Supernovae; $2$ close
binaries. In the first case one may expect GW magnitudes about
$10^{-17}\div 10^{-18}$ with the radiation in a wide frequency
range with the characteristic frequency of the order of $10^3$
$\mbox{sec}^{-1}$, in the second case magnitudes about $10^{-20}
\div 10^{-21}$ with a fixed frequency in the range of $0,1\div 10
\  \mbox{sec}^{-1}$. Due to the very small expected GW magnitudes
on the Earth, the experimental programmes intended for direct GW
detection inevitably come across the problem of noise of external
thermal and quantum nature. This struggle is already in its third
decade and requires the creation of high-precision deeply cooled
detectors.

On the other hand, it is well-known that even such weak-magnitude
GW carry rather a high energy: in the above examples, this energy
is of the order of1 W/cm2 in the first case and about
$10^{-13}\div 10^{-11}$ $\mbox{W/cm}^2$ in the second case.
Electromagnetic signal detection on such a power level has no
problems. Therefore, the GW detection problem should be solved in
a different way: by looking for specific electromagnetic signals
from GW effect upon matter in those regions of the Galaxy where
the gravitational radiation intensity is high. Setting the problem
in such a way, we should above all study the GW effect on
plasma-like media. The corresponding studies, carried in the
eighties mainly in the Kazan school of gravitation, revealed a
number of specific electromagnetic reactions of plasma to GW. In
Refs. \cite{2}–- \cite{5} the effect of plane GW (PGW) on
plasma-like media was investigated by the methods of relativistic
kinetic theory in the approximation of negligible back reaction of
matter on the PGW:
\begin{equation}\label{eq21}
(8\pi G/c^2)\varepsilon \ll \omega^2.
\end{equation}
where $\omega$ is the GW characteristic frequency, $\varepsilon$
is the matter energy density. These papers have revealed a number
of phenomena of interest, consisting in induction of longitudinal
electric oscillations in the plasma by PGW. In spite of the
strictness of the results obtained in \cite{2}--\cite{5}, the
effects discovered have very little to do with the real problem of
GW detection. Moreover, the above calculations show a lack of any
prospects for GW detectors based on dynamical excitation of
electric oscillations by gravitational radiation. There are two
reasons for that: the smallness of the ratio $m^2G/e^2 = 10^{-43}$
and the small relativistic factor $\langle v^2\rangle /c^2$ of
standard plasmalike systems. The GW energy conversion coefficient
to plasma oscillations is directly proportional to a product of
these factors.

However, the situation may change radically if strong electric or
magnetic fields are present in the plasma. In Ref. \cite{6}, where
the induction of surface currents on a metal-vacuum interface by a
PGW was studied, it was shown that the values of currents thus
induced can be of experimental interest. In \cite{7}, on the basis
of relativistic kinetic equations, a set of magneto-hydrodynamics
(MHD) equations was obtained, which described the motion of
collisionless magnetoactive plasma against the background of a PGW
of arbitrary magnitude in drift approximation and it was shown
that, provided the propagation of the PGW is transversal, there
arises a plasma drift in the PGW propagation direction.

In Ref. \cite{1} an exact solution of the relativistic MHD
equations in the PGW background of arbitrary magnitude was
obtained and, on its basis, a fundamentally new class of
sufficiently nonlinear threshold effects was discovered, named
GMSW (“jimmysway”) - gravimagnetic shock waves.

\subsection{GMSW}
The PGW metric of the polarisation e+ is described by the
expression \cite{torn}:
\begin{equation}
\label{4.1} d s^{2}=2 du dv - L^{2}[e^{2 \beta}(dx^{2})^{2} +
e^{-2\beta}( dx^{3})^{2}],
\end{equation}
where $\beta(u)$ is an arbitrary function (the PGW magnitude),
while $L(u)$ (the PGW background factor) obeys the ordinary
second-order differential equation
\begin{equation}\label{4.2}
L '' + L\dot{\beta}^2 = 0;
\end{equation}
$u = 1/ \sqrt{2}(t-x^1)$ is the retarted time and $v = 1/
\sqrt{2}(t+x^1)$ is the advanced time. Let there be no PGW at
($u\leq 0$):
\begin{equation}\label{4.3}
\left.\beta(u)\right|_{u\leq 0} = 0; \quad
\left.L(u)\right|_{u\leq 0} = 1,
\end{equation}
while the plasma be homogeneous and at rest:
\begin{equation}\label{4.4}
\begin{array}{ll}
\left.v^v(u)\right|_{u\leq 0} = \left.v^v(u)\right|_{u\leq 0}
=1/\sqrt{2}; & \left.v^2\right|_{u\leq 0} =\left.v^3\right|_{u\leq 0}
=0;\\[12pt]
\left.\varepsilon(u)\right|_{u\leq 0};\quad
\left.p(u)\right|_{u\leq0}= p_0 & \\
\end{array}
\end{equation}
($p = p(\varepsilon$) is the plasma pressure, $v^k$ is its dynamic
velocity vector) and a homogeneous magnetic field is directed in
the $(x^1, x^2)$ plane:
\begin{equation}\label{4.5}
\begin{array}{ll}
\left.H_1(u)\right|_{u\leq 0} = H_0\cos\Omega; &
\left.H_2(u)\right|_{u\leq 0} = H_0\sin\Omega;\\[12pt]
\left.H_3(u)\right|_{u\leq 0} = 0; & \left.H_\alpha(u)\right|_{u\leq 0} = 0,\\
\end{array}
\end{equation}
where $\Omega$ is the angle between the axis $Ox^1$ (the PGW
propagation direction) and the magnetic field $\mathbf{H}$
direction. The conditions (26) correspond to the vector potential
:
\begin{displaymath}
A_{v}=A_{u}=A_{2}=0;
\end{displaymath}
\begin{equation}
\label{4.6} A_{3}=H_{0} (x^{1} \sin\Omega - x^{2} \cos\Omega);
\hspace{1.5 cm} (u \leq 0).
\end{equation}
The exact solution of the relativistic MHD equations against the
metrics background (\ref{4.1}). The exact solution of the
relativistic MHD equations against the metrics background
(\ref{4.1}) obtained in \cite{1} satisfies the initial conditions
(24) - (26) and is determined by the governing function :)
obtained in \cite{1} satisfies the initial conditions (\ref{4.4})
- (\ref{4.6}) and is determined by the {\it governing function}:
\begin{equation}
\label{4.7} \Delta(u) \stackrel{Df}{=} 1 - \alpha^{2}(e^{2\beta} -
1),
\end{equation}
where $\alpha$ is a dimensionless parameter:
\begin{equation}
\label{4.8} \alpha^{2}=
 \frac{H^{2}_{0}\sin^{2}\Omega}{4\pi(\varepsilon_{0} + p_{0})}.
\end{equation}
This solution contains a physical singularity on the hypersurface
$\Sigma:\ u = u_*$:
\begin{equation}
\label{4.9} \Delta(u_*) = 1 - \alpha^{2}(u_*)(e^{2\beta(u_*)} -
1)=0,
\end{equation}
where the plasma and magnetic field energy densities tend to
infinity and the dynamic velocity of the plasma as a whole tends
to the velocity of light in the PGW propagation direction. In this
case the ratio of the magnetic field energy density and the plasma
energy tends to infinity. This singularity is a gravimagnetic
shock wave (GMSW, \cite{1}), spreading in the PGW propagation
direction at a subluminal velocity. According to Eq. (\ref{4.9}),
necessary conditions for the occurrence of the singularity are
\begin{equation}\label{4.10}
\beta(u)>0;
\end{equation}
\begin{equation}\label{4.11}
\alpha^2>1.
\end{equation}

An extrimely important fact is that a singular state is even
possible in a weak PGW ($|\beta|\ll1$) under the condition that
the plasma is highly magnetized ($\alpha^2\gg1$); in this case the
singularity condition arises according to (30) on the
hypersurfaces $u = u_*$:
\begin{equation}\label{4.12}
\beta(u_*)=1/(2\alpha^2).
\end{equation}
In particular, for a barotropic equation of state
($p=k\varepsilon$, $0\leq k< 1$)
\begin{equation}
\label{5.28} \varepsilon = \varepsilon_{0} \Lambda^{-1+ \nu} ;
\end{equation}
\begin{equation}
\label{5.29} v_{v} = \frac{1}{\sqrt{2}} L^{\nu} \Delta^{1 +
\frac{\nu}{2}} ;
\end{equation}
\begin{equation}
\label{5.30} \frac{v_{u}}{v_{v}} = \Delta^{-2} \left[
\Lambda^{-\nu} + (\Delta -1)^{2} L^{-2} e^{-2\beta} \cot^{2}\Omega
\right] ;
\end{equation}
\begin{equation}
\label{5.31} H^{2} = \frac{H^{2}_{0}}{\Lambda^{2}} = \left(
\cos^{2}\Omega + L^{2} \Lambda^{-\nu} e^{2\beta} \sin^{2}\Omega
\right) ,
\end{equation}
where
\[ \Lambda=L^2(u)\Delta(u),\quad \nu = \frac{2k}{1-k} > 0,\]
and
\[H^2=\frac{1}{2}F_{ik}F^{ik}\]
is the electromagnetic field invariant, (squared magnetic field
strength in the frame of reference comoving with the plasma).

It follows from (34) - (37) that if $\beta>0$, the plasma moves in
the GW propagation direction ($v^1 = 1/\sqrt{2}(v_u - v_v) > 0$)
and if  $\beta< 0$, in the opposite direction. The effect is
maximum in the PGW propagation direction, that is, perpendicular
to that of the original magnetic field strength, and vanishes in
the direction parallel to the magnetic field strength.

In the case of strictly transversal PGW propagation
($\Omega=\pi/2$), in the direction $Ox^2$ the plasma drift
vanishes, and the component of the plasma physical 3-velocity in
the $Ox^1$ direction $v^1$ is
\begin{equation}\label{38}
v^1=c\frac{v_u-v_v}{v_u+v_v}=c\frac{1-\Delta^2
\Lambda^\nu}{1+\Delta^2 \Lambda^\nu}.
\end{equation}
The components of the total (including the magnetic field) EMT of
the magnetoactive plasma, $T^\alpha_k$ ($\alpha=1,4$) have a
hydromagnetic structure:%
\begin{equation}\label{39}
T^\alpha_k=({\cal E}+P)v^\alpha v_k-P\delta^\alpha_k,
\end{equation}
where
\begin{equation}\label{40}
\varepsilon_H=P_H=\frac{H^2}{8\pi};\quad {\cal
E}=\varepsilon+\varepsilon_H;\quad P=p+P_H,
\end{equation}
where $P$ and ${\cal E}$ are the total pressure and energy density
of the magnetoactive plasma. There arises an energy flow in the
plasma in the direction $Ox^1$:
\begin{equation}\label{41}
T^{14}=\frac{\varepsilon_0+p_0}{4L^4}(\Delta^{-4}\Lambda^{2\nu}-1)(\Delta\Lambda^{2\nu}+\alpha^2
e^{2\beta}).
\end{equation}
The parameter $\nu$ takes in these formulae the following values
in the two extreme cases:
\begin{equation}\label{42}
\nu=\left\{\begin{array}{ll} 0; & k=0;\\
1; & k=1/3.
\end{array}\right.
\end{equation}
For a weak GW
\begin{equation}\label{43}
|\beta(u)|\ll 1; \quad L^2(u)=1+O(\beta^2)\approx 1,
\end{equation}
the expressions (37), (41) and (38) take the form
\[ \frac{v^1}{c}=\frac{1-\Delta^m}{1+\Delta^m};
\]
\begin{equation}\label{44}
T^{14}=\frac{1}{4}(\varepsilon_0+p_0)(1+\alpha^2)(\Delta^{-n}-1);
\end{equation}
\begin{equation}\label{45}
\frac{H^2}{H^2_0}=\frac{1}{\Delta^m(u)};
\end{equation}
where the coefficients $m$ and $n$ take integer values for
nonrelativistic ($k = 0$) and ultrarelativistic ($k = 1/3$)
plasma:
\begin{equation}\label{46}
\begin{array}{lll}
k=0; & m=2; & n=4;\\
k=1/3; & m=3; & n=2.\\
\end{array}
\end{equation}

\section{GW energy transmission to plasma: a
half-self-consistent solution}
\subsection{Total momentum conservation}
Since on the singular hypersurface (30) $\Delta(u)=0$, the energy
densities of the plasma and the magnetic field tend to infinity,
and the velocity of the plasma as a whole tends to the speed of
light, the total energy of the magnetohydrodynamic shock wave and
its flow in the GW propagation direction tend to infinity. The
singular state emerging in the plasma due to the PGW violates the
basic assumption (21) of the weaknesss of GW interaction with the
plasma. In a more complete self-consistent problem including
gravitation, the back reaction of the shock wave upon the PGW
should lead to PGW energy loss and its magnitude damping up to the
values
\begin{equation}\label{47}
{\mbox max}|\beta|<1/2\alpha^2.
\end{equation}
Thus a GMSW is an effective mechanism of a gravitational wave
energy pumping over into plasma \cite{1}. A rigorous solution of
the problem of PGW energy transformation into the shock wave
energy is only possible by studying the self-consistent set of the
Einstein equations and the MHD equations.

Ref. \cite{1} suggested a semiquantitative solution of this
problem on the basis ofa simple model of energo--ballance. Due to
its extreme importance, we do not restrict ourselves to \cite{1}
and return to a more complete study of the problem of energy
transmission from a GW to magnetoactive plasma. However, instead
of solving the Einstein equations, we make use of their
consequence, the conservation law of the total momentum of the
system “plasma + gravitational waves”. Clearly, this model is only
approximate and cannot replace a rigorous solution to the Einstein
equations. According to \cite{land}, an arbitrary gravitational
field provides the conservation of the system’s total momentum
\begin{equation}\label{48}
P^i=\frac{1}{c}\int(-g)(T^{i4}+t^{i4})dV
\end{equation}
where $t^{ik}$ is the energy-momentum pseudotensor of the
gravitational field and the integration covers the whole
3-dimensional space. Let us take into account that the above
solution is plane-symmetric and only depends on the retarded
“time” $u$. Consequently the integration over the “plane”
$(x2,x3)$ in (\ref{48}) reduces to simply multiplying by an
infinite 2-dimensional area. Dividing both sides of (\ref{48}) by
this area and bearing in mind that with $\Omega=\pi/2$ among the
3-dimensional flows only $P^1$ is nonzero, we obtain the
conservation law of the surface density of the momentum
$P^1_\Sigma$:
\begin{equation}\label{49}
P^1_\Sigma=\frac{1}{c}\int\limits_{-\infty}^{+\infty}(-g)(T^{i4}+t^{i4})x=\mbox{Const}.
\end{equation}
Let the right semispace $x > 0$ be filled with magnetoactive
plasma and the left one $x < 0$ with matter which does not
interact with a weak GW. Let further the whole gravitational
momentum be concentrated in the interval $u\in[0,u_f]$, where $t_f
= \sqrt{2}u_f$ is the gravitational pulse duration. Since the
integral in Eq. (49) is conserved all the time, let us consider it
at $t_0 < 0$, when the GW has not yet reached the magnetoactive
plasma, and $t_f > t > 0$, when the GW has reached the plasma.
Taking into account that the vacuum solution depends only on the
retarded time, we get for the integral in Eq. (49):
\begin{equation}\label{50}
\int\limits_0^{u_f}t^{14}_0=\int\limits_0^{t/\sqrt{2}}(T^{14}+t^{14})du+\int\limits_{t/\sqrt{2}}^{u_f}t^{14}_0,
\end{equation}
where $t^{14}_0 = t^{14}(\beta_0(u))$; $t^{14} =
t^{14}(\beta(u))$, $\beta_0(u)$ is the vacuum magnitude of the
PGW, $\beta(u)$ is the PGW magnitude with allowance for
interaction with the plasma. Transferring one of the integrals to
the left-hand side of Eq. (50), we arrive at the relation
\begin{equation}\label{51}
\int\limits_0^u t^{14}_0=\int\limits_0^u(T^{14}+t^{14})du,
\end{equation}
where the variable $u = t/\sqrt{2} > 0$ can now take {\it any
positive} values.

A similar law may be written for the plasma total energy; in this
case instead of Eq. (51) we obtain:
\[
\int\limits_0^u t^{44}_0=\int\limits_0^u(T^{44}-{\cal
E}_0+t^{14})du,
\]
where ${\cal E}_0$ is the total energy density of the unperturbed
plasma.

\subsection{Local analysis of the conservation law}
Since the relation (51) must be valid at any values of the
variable $u$, the corresponding local relation should hold:
\begin{equation}\label{52}
T^{41}(\beta) + t^{41}(\beta) = t^{41}(\beta_0),
\end{equation}
i.e. a local conservation law of the energy flow density should
hold, as was assumed in Ref. \cite{1}. It should be pointed out
that the local conservation law (52) is a direct consequence of
the solution stationarity, i.e. the solution dependence on the
retarted time $u = (ct - x)/ \sqrt{2}$. There are two factors
preventing the solution in a rigorous model from being stationary:
(1) PGW interaction with the plasma; (2) the boundary conditions
on the surface $x = 0$. In accordance with the approximation (21),
we introduce a small dimensionless parameter $\chi$ \cite{1}:
\begin{equation}\label{53}
\chi^2=\frac{\pi G
(\varepsilon_0+p_0)(1+\alpha^2)}{c^2\omega^2}\sim\frac{\omega^2_g}{\omega^2},
\end{equation}
where $\omega$ is the characteristic GW frequency,
\[\omega^2_g = \frac{8\pi G{\cal E}_0}{c^2}. \]
The approximation (21) is equivalent to the condition
\begin{equation}\label{54}
\chi^2\ll1.
\end{equation}
Under the condition (54) the GW velocity tends to that of light,
thus providing the required solution stationarity even in
inhomogeneous plasma \cite{24}. Let $\beta_0=\mbox{ Const}> 0$ be
a maximum value of the PGW vacuum magnitude, $\beta_*$. Let us
introduce one more dimensionless parameter, the first GMSW
parameter $\xi^2$:
\begin{equation}\label{55}
\xi^2=\chi^2/\beta^2_0\sim {\cal E}/{\cal E}_{GW}
\end{equation}
where
\[{\cal E}_{GW}=\beta^2_0\omega^2c^2/(4\pi G).\]
Thus, the parameter $\xi^2$ is of the order oft he ratio of the
total magnetoactive plasma energy to the vacuum GW energy.

Making use of the solution of the MHD equations in the case of
strict transversal PGW propagation ($\Omega= pi/2$) as well as the
expression of the total plasma EMT (44) and that for the energy
flow of a weak PGW (3), we reduce Eq. (52) to the form
\begin{equation}\label{56}
\dot{\beta}^2+\overline{V}(\beta)=\beta^2_*,
\end{equation}
where
\begin{equation}\label{57}
\overline{V}(\beta)=\chi^2[\Delta^{-n}(\beta)-1] \quad (\nu=1)
\end{equation}
is a function of $\beta$; $\dot{\beta}$ is now a derivative in the
dimensionless ``time'' $s=\sqrt{2}\omega u/c$.

Introduce the relative PGW magnitude:
\begin{equation}\label{58}
q=\beta/\beta_0; \quad q_*=\beta_*/\beta_0.
\end{equation}

Then Eq. (56) may be rewritten in the form
\begin{equation}\label{59}
\dot{q}^2+V(q)=\dot{q}^2_*,
\end{equation}
where
\begin{equation}\label{60}
V(q)=\xi^2[(1-\Upsilon q)^{-n}-1]
\end{equation}
and a new dimensionless parameter has been introduced:
\begin{equation}\label{61}
\Upsilon=2\alpha^2\beta^2_0
\end{equation}
(the second GMSW parameter). The total energy conservation law
leads roughly to the same result. Eq. (59) may be treated as an
equation with respect to the variable $q$ . On the other hand,
(59) completely coincides in its form with the energy conservation
law of a 1-dimensional mechanical system described by the
canonical variables $\{q(s),\dot{q}(s)\}$ [23], where $V(q)$ is
the potential, $\dot{q}^2$ is its kinetic energy and
$\dot{q}^2_*=E_0$ is its total energy. Fig. 4. shows the
qualitative form of the potential $V (q)$.

\ris{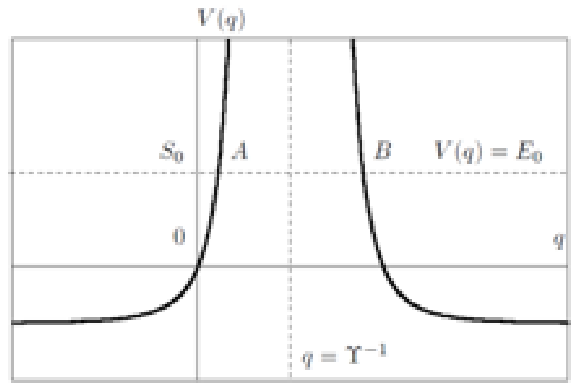}{Potential $V(q)$ of Eq.(59)}

Two points on the potential curve (A and B) correspond to any
positive value of $E_0$. These points are the system trajectory
turning points. No real system states exist under the potential
curve $V(q)$. At the point
\begin{equation}  \label{62}
q = q_c = \Upsilon^{-1}; \quad (\beta = 1/2\alpha^2)
\end{equation}
\begin{equation}  \label{63}
V(q_c)\to \infty.
\end{equation}

To analyze the system behaviour, let us suppose that the moment
$s=0$ corresponds to the front edge of the GW, while
\begin{equation}  \label{64}
\beta_*\approx \beta_0\sin{s}\Rightarrow q_*\approx \sin(s).
\end{equation}
Thus, provided the initial conditions (24) are satisfied, the
system always starts from the point $S_0$ along the line $(AB)$
towards $A$ (for $\beta > 0$). Since this is a turning point, the
maximum accessible value of the variable $q$ in the system is
$q(A)$. This is the smallest root $q_{-}=q(\chi,\Upsilon,E_0)$ of
the algebraic equation
\begin{equation}  \label{65}
V(q) - E_0 = 0.
\end{equation}
The maximum attainable PGW magnitude in the system, $\beta_{max}$,
is
\begin{equation}  \label{66}
\beta_{max} = q_{-}\beta_0.
\end{equation}
Thus $q_{-}$ coincides in its sense with the “PGW magnitude
damping factor” $\gamma$ introduced Ref. \cite{1}. Solving Eq.
(\ref{65}), we obtain the required root $q_{-}$:
\begin{equation}  \label{67}
q_{-} = \frac{1}{\Upsilon}\left[1 - \left(1 +
\frac{\dot{q}^2_*}{\xi^2}\right)^{-1/n}\right].
\end{equation}
From (\ref{65}) it follows that always
\begin{equation}  \label{68}
q_{-} \leq \Upsilon^{-1},
\end{equation}
and also, as $E_0\to 0$
\begin{equation}  \label{69}
q_{-}\approx \frac{\dot{q}^2_*}{4\Upsilon \xi^2}\to 0.
\end{equation}
With increasing $E_0$ this magnitude grows and for $E_0\to \infty$
it reaches the value  $(q = q_c )$:
\begin{equation}  \label{70}
\beta_{max}\to \frac{\beta_0}{\Upsilon}.
\end{equation}
After the turning point the GW magnitude diminishies, reaching
negative values. For $s\to+\infty$
\begin{displaymath}
s\to +\infty \quad \beta' \to \beta'_{\infty} = \const < 0; \quad
\beta \sim \beta'_{\infty} u \to -\infty ;
\end{displaymath}
the metric (22) degenerates $(g_{22}\to 0, g_{33}\to -\infty)$;
the only nonzero components of the curvature tensor take the
following form due to the Einstein equations (see (23)):
\begin{displaymath}
R_{u2u2} = (L^2)'\beta'_{\infty}\exp(2\beta'_{\infty}u)\to -0;
\end{displaymath}
\begin{displaymath}
R_{u3u3} = (L^2)'\beta'_{\infty}\exp(2\beta'_{\infty}u)\to +\infty
.
\end{displaymath}
Thus, as $s\to +\infty$, a true singularity is formed in the
system. It is easily verified that in this case $H^2\to 0$,
$\varepsilon \to 0$, $V^1\to -c$. The plasma in the final state
moves to meet the original GW direction. this reverse of the
plasma needs a more detailed self-consistent analysis.

\subsection{Numerical analysis ofGMSW}
Let us pass to a more detailed study on the selfconsistent motion
of the system. From Eq. (59) we obtain the differential equation
\begin{equation}\label{71}
dq/ds=\pm \sqrt{q^2_*-V(q)}
\end{equation}
where the plus sign is chosen before and the minus sign after the
turning point $q_-$. It is helpful to solve and analyze Eq. (71)
using the new dimensionless variables: $\Delta(\beta)$ and
\begin{equation}\label{72}
S=\Upsilon s\equiv \sqrt{2}\Upsilon \omega u
\end{equation}
Substituting into (71), for example, $q_*(s) = \sin s$, we reduce
it to the form
\begin{equation}\label{73}
d\Delta/dS=\mp \sqrt{\cos^2S/\Upsilon-V(\Delta)},
\end{equation}
moreover, the initial condition is to be fulfilled:
\begin{equation}\label{74}
\Delta(0)=1.
\end{equation}

Figs. 5–11 show some results of numerical integration of Eq. (73)
with the initial condition (74). An analysis of the formulae
describing GMSW and numerical calculations make it possible to
discover a number of general laws of the GMSW excitation process
in homogeneous and isotropic plasma under the condition that the
PGW propagation is strictly transversal:

\begin{enumerate}
\item A GMSW is completely described by three nonnegative
dimensionless parameters: the parameter $k$ in the plasma equation
of state, the first ($\xi^2$) and the second ($\Upsilon$) GMSW
parameters.
\item Necessary conditions for GMSW excitation are
(31) and (32):
\begin{equation}\label{75} \Upsilon\geq 1.
\end{equation}
\item The only criterion of strong GW absorption is,
according to (69), a large value of the second GMSW parameter
($\Upsilon$):
\begin{equation}\label{76} \Upsilon\gg 1.
\end{equation}
\item Under these conditions a maximum response of the plasma to GW\footnote{Here and further on, speaking of a
maximum response of the plasma, we mean its energy
characteristics: the plasma energy flow density and the magnetic
field energy density.} is achieved when the values of the first
GMSW parameter are small:
\begin{equation}\label{77} \xi^2\ll 1.
\end{equation}
\item The plasma response to GW is a single pulse,
and the shock wave stage is always replaced by a reverse stage,
when the plasma turns back. Simultaneously its density, pressure
and magnetic field strength fall off.
\item The ultrarelativistic ($k = 1/3$) plasma response
is much greater (by approximately 2 orders) than that of a plasma
with the nonrelativistic equation of state ($k = 0$), and in
ultrarelativistic plasma the pulse duration is also slightly
greater.
\item The profiles of the plasma response at sufficiently
large values of the second GMSW parameter ($\Upsilon\geq5$)
actually coincide on the $S$ scale. This means that in the
conventional time scale $t$ the pulse duration is inversly
proportional to the second parameter, or more precisely
\begin{equation}\label{78}
\Delta\tau\approx
\frac{\pi}{2\omega\Upsilon}=\frac{T}{4\pi\Upsilon},\quad
(\Upsilon\geq 5),
\end{equation}
where $T$ is the GW period.
\item With $\Upsilon< 5$ the response magnitude rapidly decreases
and at $\Upsilon\sim 1$ becomes smaller by an order of magnitude.
In this case a maximum pulse duration is achieved:
\begin{equation}\label{79}
\tau\leq T/4=\pi/(2\omega).
\end{equation}
\item A decrease in the first GMSW parameter causes
a rapid increases in the response (roughly propotional to
$1/\xi^2$); simultaneously increases the pulse duration
approximately by a factor of 2.
\item A maximum response is achieved at the instant
$S\approx 1$.
\item Under the optimal GMSW conditions (76) and
(77), the total surface density of the magnetic field energy,
transported in the pulse, ${\cal E}_\Sigma$, is f or
ultrarelativistic plasma of the order of
\begin{equation}\label{80}
{\cal E}_\Sigma\sim cH^2_0/(\omega\xi^2\Upsilon).
\end{equation}
\item The shock wave energy is taken from the GW
energy, so under the conditions (76) and (77) the GMSW is an
effective mechanism of gravitational waves energy transformation
into other forms of energy.
\end{enumerate}

\ris{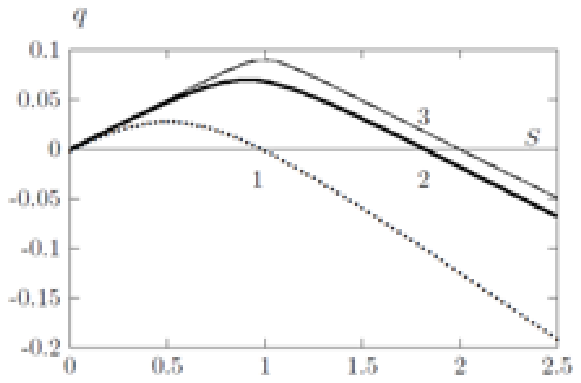}{Relative GW magnitude in ultrarelativistic plasma,
$q(S)$; everywhere $\Upsilon = 10$. 1 —- $\xi^2 = 1$; 2 —
$\xi^2=0,1$; 3 —- $\xi^2 = 0,01$.}

\ris{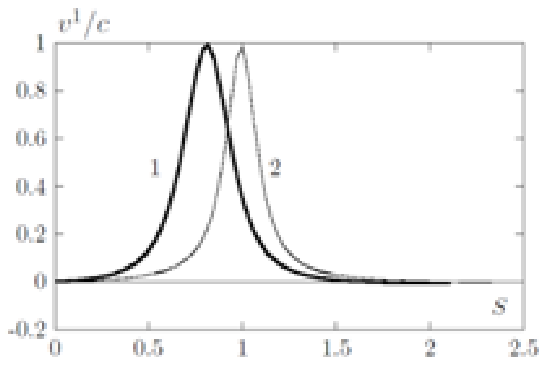}{Physical velocity of plasma in GW
field,$v^1(S)/c$: 1 —- nonrelativistic, 2 —- ultrarelativistic
equation of state; $\Upsilon$ = 10; $\xi^2 = 0, 01$. }

\ris{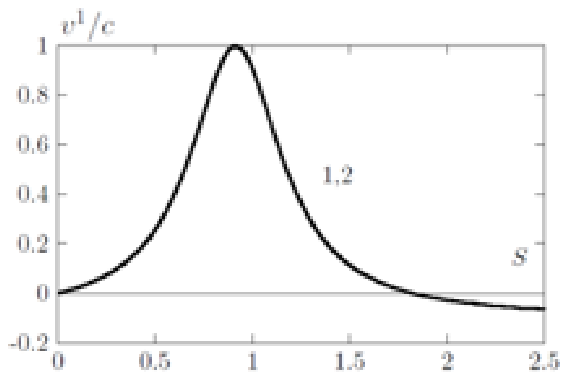}{Physical velocity of plasma in GW
field,$v^1(S)/c$: 1 -- $\Upsilon$ = 10; 2 -- $\Upsilon=100$.
Everywhere $\xi^2 = 0,1$. The lines practically coincide.}

\ris{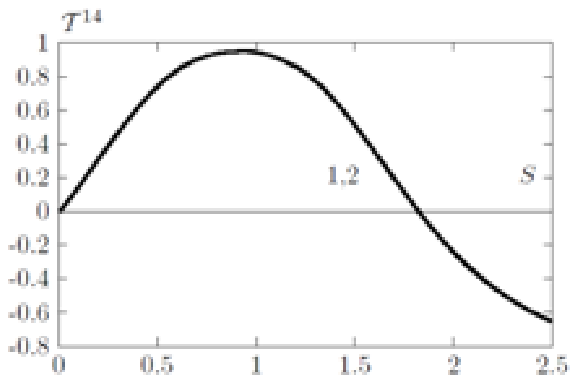}{Dimensionless density of ultrarelativistic plasma
energy flow, ${\cal T}^{14}(S) = 4\pi G T^{14}/\beta^2_0 \omega^2
c^2$: 1 —- $\Upsilon$ = 10 ;2 —- $\Upsilon$ = 100. Everywhere
$\xi^2=0,1$. The lines practically coincide.}

\ris{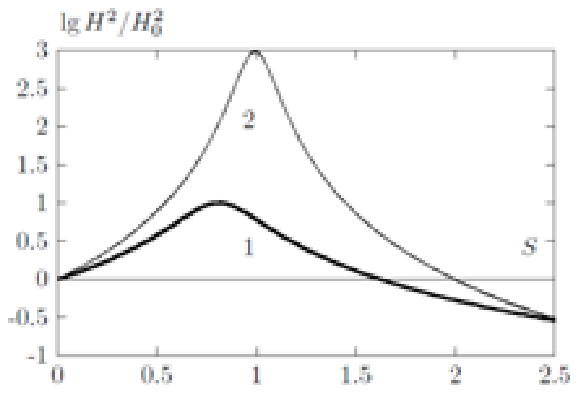}{Magnetic field strength $\lg H^2(S)/H^2_0$ in GW
field, 1 —- $k = 0$ ,2 —- $k = 1/3$; $\Upsilon = 10$; $\xi^2 =
0,01.$.}

\ris{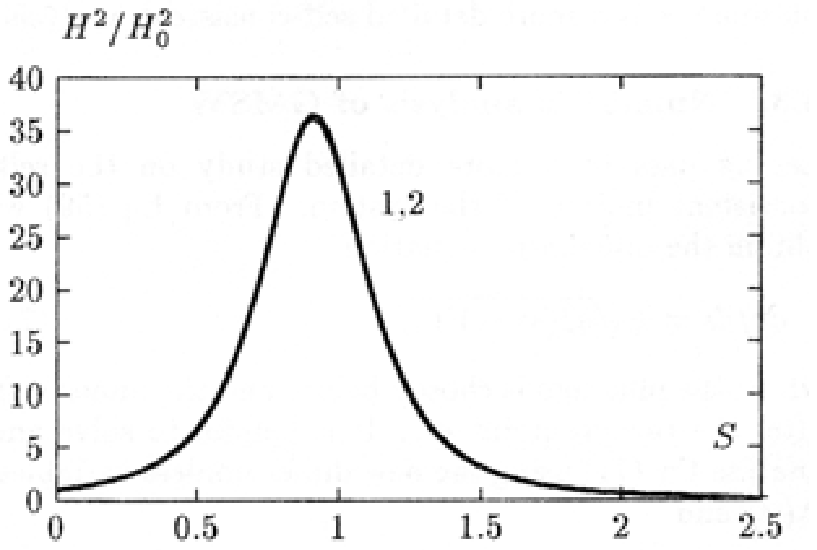}{Magnetic field strength in GW field for
ultrarelativistic plasma, $H^2(S)/H^2_0$: 1 —- $\Upsilon = 10$; 2
—- $\Upsilon = 100$. Everywhere $\xi^2 =0,1$. The lines
practically coincide.}

\ris{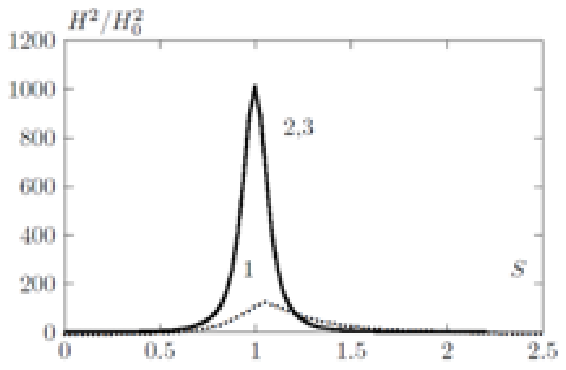}{Magnetic field strength in GW field for
ultrarelativistic plasma, $\lg H^2(S)/H^2_0$: 1 —- $\xi^2 = 1$; 2
—- $\xi^2 = 0,1$; 3 —- $\xi^2 = 0,01$. Everywhere $\Upsilon =
10.$}

\ris{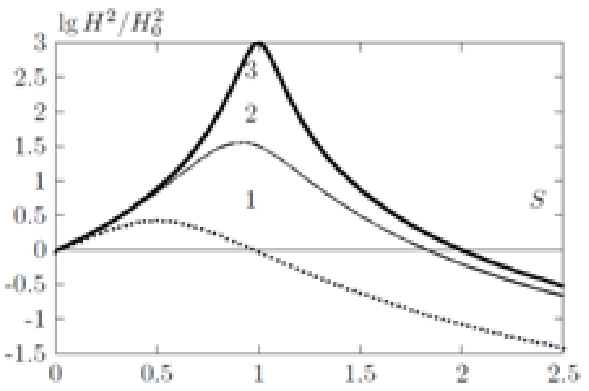}{Magnetic field strength in GW field for
ultrarelativistic plasma, $H^2(S)/H^2_0$: 1 —- $\Upsilon = 1$; 2
——- $\Upsilon = 10$; 3 ——- $\Upsilon = 100$. Everywhere $\xi^2
=0,01$. The lines 2 and 3 practically coincide.}

\section{GMSW in neutron star magnetospheres}
\subsection{GSMW parameters in neutron star
magnetospheres}
In [1] it was shown that in the magnetospheres ofneutron stars
performing quadrupole oscillations, large values of the second
GMSW parameter are realized. That is, the necessary condition for
GMSW excitation (76) is fulfilled. Let us study this problem in
more detail. The electron number density in a pulsar
magnetosphere, $n_e(r)$, which is necessary for calculating the
parameter $\Upsilon$, may be obtained by dimensional estimation
from the Maxwell equations [12]:
\begin{equation}\label{81}
n_e(r)\sim H(r)/(4\pi er).
\end{equation}
Further, as is known from (see \cite{15}), the pulsar period
slowing-down rate $t_0$ is connected with the pulsar parameters as
follows:
\begin{equation}  \label{GSMW636}
t_0\approx \frac{3c^2MP^2}{8\pi^2 H^2R^4},
\end{equation}
where $R$ is the neutron star radius, $M$ is its mass, $H$ is the
magnetic field strength at the stellar surface, $P$ is the
rotation period. This formula gives for the pulsar NP 0532 $NP \,
\, 0532 \, \, H\approx 5\cdot 10^{12}$ G (see Table 3) according
to the known slowing-down rate of this pulsar. Actually, as
pointed out in Ref. \cite{11}, the magnetic field strength at the
surface of NP 0532 is somewhat smaller than the value obtained on
the basis of (82), and is of the order of $10^{12}$ G. Further on
we use this value. In the figures shown below (unless specially
indicated) the following values of the parameters are adopted: $R
= 1,2\cdot 10^6$ cm, $\beta_0(R) = 10^{-8}$ and the magnetic field
in the magnetosphere is assumed to be dipole: $H(r)\sim (R/r)^3$.

\vskip 12pt \noindent \refstepcounter{table}Table \thetable. {\sl
GMSW in a neutron star magnetosphere}\\[8pt]
{\small
\begin{tabular}{|c|c|c|c|c|c|c|c|c|c|c|c|}
\hline
 & & & & & & & & & & & \\[-4pt]
 $r_g/R$ & $\delta M$& $T_n$ & $\tau_n$& $Em/\Delta_R$ &
$L_g/\Delta_R$ & $R$ & $H(R)$ & $\sqrt{\Delta_R}$ & $\alpha^2$ &
$\Upsilon$ & $E_m$\\[4pt]
\hline
 & & & & & & & & & & & \\[-8pt]
0,057 & 0,405 & 1,197 & 13,0 & 7,8(50) & 1,2(50) & 21 &
7,7(11)& 1,0(-4)& 5,2(11) & 1,1(5)& 8,4(42)\\
\hline
 & & & & & & & & & & & \\[-8pt]
0,159 & 0,677 & 0,699 & 1,7 & 5,7(52) & 7,0(52) & 13 & 2,7(12)& 4,2(-6)& 4,7(11)& 2,4(4)& 1,1(42)\\
\hline
0,240 & 0,682 & 0,311 & 0,2 & 2,8(52)& 2,9(53)& 13 & 6,1(12)& 3,3(-6)& 7,5(11)& 4,4(4) & 3,0(41)\\
\hline
0,580 & 1,954 & 0,378 & 0,2 & 1,7(54) & 1,6(55) & 10 & 7,4(12)& 4,2(-7)& 8,0(11)& 4,2(4)& 3,1(41)\\
\hline
0,434 & 1,670 & 0,349 & 0,2 & 5,0(53)& 5,0(54) & 12 & 5,2(12)& 9,5(-7)& 6,3(11)& 4,1(4) & 4,7(41)\\
\hline
\end{tabular}}
\vskip 8pt\noindent $^*$ Comments to Table 3. The data placed in
columns 1 - 7, 9 and 12 are taken from the book \cite{torn}.
$\delta M =M/M_\odot$ is the neutron star mass related to the
Solar mass; $R$ is the star radius in km; $T_n$ is the neutron
star eigen-oscillation period in the basic quadrupole mode (in
milliseconds); $\tau_n$ is the oscillation damping time (in
seconds); $\Delta R=\langle(\delta R/R)^2 \rangle$ is the
root-mean-square relative magnitude of the neutron star
oscillations; $E_m$ is the oscillations kinetic energy in erg;
$L_g$ is the the star's gravitational luminosity in erg/sec. The
GW magnitude value at the neutron star surface, $\beta_0(R)$, is
assumed to be equal to $10^{-8}$. $H(R)$ is the magnetic field
strength in gauss (G). The data placed in columns $8\div$11 are
calculated using Eqs. (3), (82) and (81) for the observed Crab
pulsar (NP 0532) parameters; $P$ = 0,033 s, $t_0$ = 2500 years.
The data placed in the last line of the table (columns 2 $\div$ 7
and 9) are obtained by extrapolation of the values from the book
\cite{torn}. The values of the parameters $\alpha^2$ and
$\Upsilon$ are given for the magnetosphere near the stellar
surface. \vskip 8pt \hrule \vskip 8 pt

If the magnetic field of a neutron star is described as that of a
dipole, then the geographic angle $\Theta$ (counted from the
magnetic equator) will be connected with the above angle $\Omega$
by the relation $\Omega = \pi /2 - \Theta$. Therefore the GMSW
excitation condition depends on the angle $\Theta$:
\begin{displaymath}
\sin^2\Theta < 1 - \frac{1}{2\alpha^2_0|\beta|}\sim 1 -
\Upsilon^{-1}.
\end{displaymath}

Thus, in the magnetosphere of a neutron star (or a Supernova) a
GMSW can be excited in the vicinity of the magnetic equator,
similarly to pulsars, with a knife radiation pattern. In this
region, as was demonstrated by the above examples, the
gravitational radiation can be absorbed almost completely by shock
wave excitation. Fig. 13 shows the radial dependence of the GMSW
parameters magnetic equator plane ($\Theta=0$) of a neutron star
magnetosphere with the above parameters $R$, $H(R)$ and
$\beta_0(R)$\footnote{In all further figures the magnetosphere is
considered in the magnetic equator plane.}. According to Table 3,
in the case of NP 0532 such values of the parameters $\beta_0(R)$
and $R$ correspond to the gravitational radiation power
$L_g\approx 4,5 \cdot 10^{42}$ erg/sec.

\ris{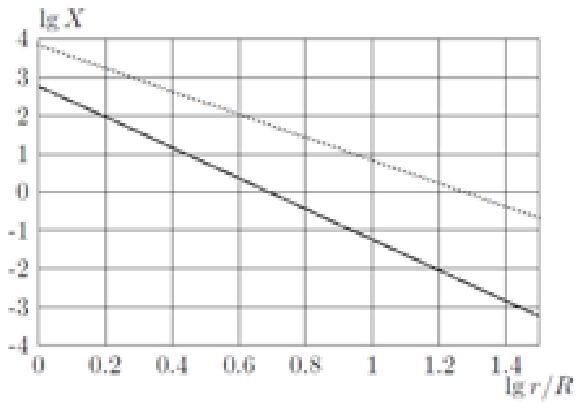}{GMSW parameters in a neutron star magnetosphere
for a dipole radial dependence of the magnetic field strength. The
line corresponds to $\lg \xi^2$; the points correspond to
$\lg\Upsilon$.}

As is seen from Fig. 13, the region favourable for the GMSW
formation lies in the range $6R\div 16R$, i.e., where the local
magnetic field strength is $3·10^8\div4·10^9$ G. When the GMSW
pulse passes, these local values increase by a factor of 10 to 30.
Thus a neutron star in whose magnetosphere a GMSW zone is formed,
is able to radiate GW only from its magnetic poles, like pulsars
with a pencil radiation pattern. In this case the probability of
direct GW detection from such sources is drastically decreased.
However, GMSW open another way for GW observation. A formed GMSW
carries above all strong magnetic fields. They move from the
neutron star in the magnetic equator plane, and therefore should
lead to an increased pulsar magnetic bremsstrahlung intensity at
the moment when the GMSW front passes. Thus anomalous
electromagnetic radiation flashes in the pulsar radiation should
be observed at moments when quadrupole oscillations are excited.

The total magnetic bremsstrahlung intensity of a relativistic
electron is proportional to squared magnetic field strength
\cite{land}:
\begin{equation}\label{83}
I = 2e^4 H^2 \vec{p}^2/(3m^4c^5),
\end{equation}
where
\[\vec{p} = m\vec{v}\sqrt{1 -\vec{v}\ ^2/c^2}\]
is the electron momentum.

Therefore the curves $H^2(S)$ shown in Figs. 9–12, actually
describe the time dependence of the magnetosphere magnetic
bremsstrahlung intensity, i.e. the local electromagnetic response
to the gravitational radiation of the neutron star. Such a
response might be detected by an observer at rest placed in the
magnetosphere and screened from the electromagnetic radiation
coming from other regions. The situation is more difficult with a
total response of the magnetosphere to the GW, detected by a
distant observer. We will later return to this problem.

As was mentioned above, the response of a homogeneous
magnetoactive plasma even to strictly periodic gravitational
radiation has the form of a single pulse. But even if it were not
the case, the response of a neutron star magnetosphere to a GW
would still have the same form. Indeed, a shock wave (GMSW),
emerging after the excitation of quadrupole oscillations of a
neutron star, should throw the equatorial sector of the
magnetosphere away into the interstellar space. For the next pulse
to be formed, the magnetosphere should restore. The necessary time
for its restoration is of the order of $\Delta t \sim l/v_s$ where
$l$ is the characteristic size of the magnetosphere and $v_s$ is
the velocity of sound. For a typical neutron star magnetosphere
$\Delta t\sim 1$ sec. And typical quadrupole oscillation damping
times comprise tenths of a second, according to Table 3.

\subsection{Effect of magnetosphere inhomogeneity upon
GMSW}
According to Table 3, neutron star eigen-oscillation periods vary
depending on the stellar mass in the range of 0,3 to 1,2 ms.
Therefore the local duration of GMSW pulses should, by (79),
satisfy the condition
\begin{equation}\label{84}
\Delta\tau < 7\cdot10^{-5} \div 3\cdot 10^{-4}s, (84)
\end{equation}
i.e., be shorter than 70 to 300 microseconds. Fig. 14 shows the
dependence of the GMSW pulse local duration on the radial
coordinate $r,\Delta\tau(r)$, calculated according to Eq. (78).

\ris{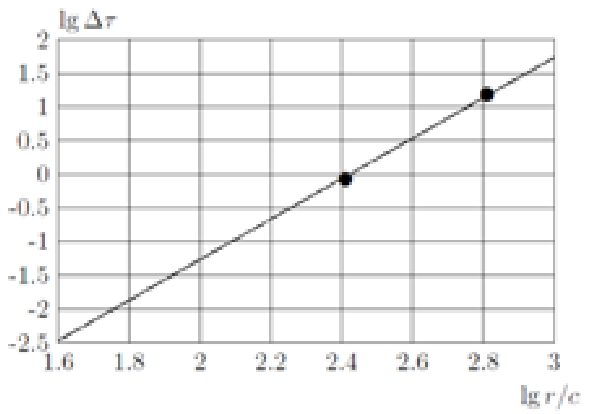}{The dependence of local pulse duration
$\Delta\tau$ (in microseconds) on the distance form the star
centre $r/c$ (in microseconds). The duration was calculated by Eq.
(78). The black circles mark the boundaries of a region where the
GMSW effect is sufficiently well-developed. Outside this region
the result is of a formal nature. On the left boundary of the
region the pulse duration should quickly grow up to the values
(84).}

As is seen from Fig. 14, the actual local pulse duration in the
region where the GMSW mechanism is fairly effective, ranges from 1
to 10 $\mu$s. Thus in this region $\Delta\tau \sim 10^{-2}r$ the
following condition is fulfilled with a large spare:
\begin{equation}\label{85}
\Delta\tau c \ll r,
\end{equation}
justifying the use of the GMSW formulae for describing an
inhomogeneous magnetosphere.

An observer out of the neutron star magnetosphere would detect the
magnetic bremsstrahlung from the magnetospheric electrons during
the whole time while the local pulse passes through the
magnetospheric region $r_- < r < r_+$ where favourable conditions
for GMSW development are realized, namely, (76), (77). With a
certain caution these conditions may be specified: $\xi^2 < 0,5$
(the lower bound of the GMSW range, $r_-$ and $\Upsilon > 5$ (its
upper bound, $r_+$). So the size of the GMSW zone is
\[\Delta r = r_+-r_- .\]

If $\Delta r < 0$, a GMSW zone does not appear in the neutron star
magnetosphere at all. Since, as we have seen, the GMSW pulse
spreading velocity is very close to $c$, the whole magnetic
bremsstrahlung detected by remote observer will be concentrated in
the time “window” of duration $\Delta T$
\begin{equation}\label{86}
\Delta T = \Delta r/c = t_+ - t_-.
\end{equation}
where $t\pm = r\pm/c$ are the instants when the GW leading front
reaches the upper and lower boundaries of the GMSW zone. Near its
boundaries $r_-$ and $r_+$ the GMSW is poorly developed (in the
first case the first GMSW parameter is too large, in the second
case the second parameter is too small). Therefore the intensity
of the electromagnetic signal is small near the boundaries of the
window, while in its medium domain a radiation maximum (large
$\Upsilon$ and small $\xi^2$) is achieved. The form of the signal
itself is yet to be calculated. Fig. 15 shows the dependence of
the window width on the magnetic field strength and the GW
magnitude.

\ris{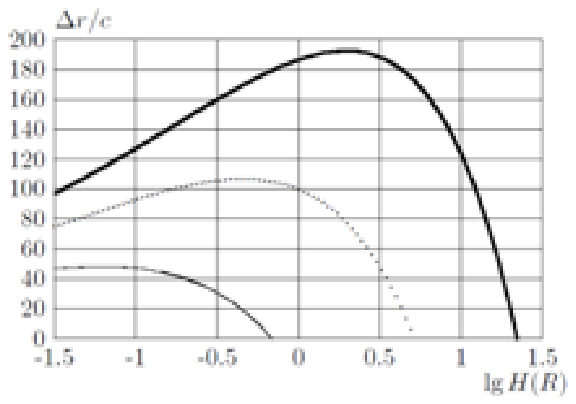}{Dependence of the GMSW existence range $\Delta T$
(in $\mu s$) in the magnetosphere of a neutron star of radius $R =
1,2 \cdot 10^6$ cm on the magnetic field strength $H(R)$, (related
to $10^{12}$ G) and the GW magnitude $\beta_0(R)$. The thin line
corresponds to $\beta_0(R) = 5 \cdot 10^{-9}$ , the points to
$\beta_0(R) = 7, 5 \cdot 10^{-9}$, the thick line to $\beta_0(R) =
10^{-8}$.}

\subsection{Magnetic bremsstrahlung intensity}
Note that the local density of the bremsstrahlung intensity
$W(t,r)$ is determined by the local values of the squared magnetic
field strength, $H^2(t, r)$, and the local electron number density
in the magnetosphere, $n_e(t, r)$. For ultrarelativistic electrons
by \cite{land} it is
\[W = \frac{2e^4H^2}{3m^2c^3}\left(\frac{E}{mc^2}\right)^2n_e,\]
where $E$ is the electron energy. Further we will assume that the
size of a local pulse is much smaller than both the characteristic
scale of the magnetospheric inhomogeneity $r$ and the window width
$\Delta r$. Thus the retarded time u is a quick variable and the
radial coordinate $r$ is a slow variable. Then a GMSW may be
described by the formulae for homogeneous plasma, where it is
necessary to use the local values of the GMSW parameters,
$\xi^2(r)$ and $\Upsilon(r)$. Mean while in the exact stationary
solutions there arises a weak dependence on the radial coordinate
$r$ , i.e. the solution will be weakly nonstationary and the
nonstationarity will show it selfin the form of a functional
dependence of the GMSW solutions on the local values of the
parameters, e.g.,
\[\Delta(u;r)=1-\Upsilon(r) q(u;r),\]
etc. Thus, from the particle number conservation law and the
solution stationarity it follows:
\begin{equation}  \label{87}
L^2n_e(r,t)v_v(u) = \const \approx \frac{1}{\sqrt{2}}n^0_e(r),
\end{equation}
where $n^o_e(r)$ is the unperturbed electron number density in the
magnetosphere. Taking into account that $En_e = \varepsilon/2$
(half energy of a relativistic magnetosphere belongs to the
electrons) and using the solutions (34), (35) and (45) for a weak
GW, we obtain:
\begin{equation}  \label{88}
n_e(r,t) = n^0_e (r)\Delta^{-3/2}(u);
\end{equation}
\begin{equation}  \label{89}
E(r,t) = E_0 (r)\Delta^{-1/2}(u);
\end{equation}
where $E_0(r)$ is the unperturbed energy of magnetospheric
electrons and
\begin{equation}  \label{90}
H^2(u) = H^2_0(r)\Delta^{-3}(u).
\end{equation}
Thus we get the following relation for the magnetic bremsstrahlung
of the ultrarelativistic magne\-to\-sphere:
\begin{equation}  \label{91}
W(r,t) = W_0 (r)\Delta^{-11/2}(u),
\end{equation}
where $W_0(r)$ is the magnetic bremsstrahlung intensity density
for an nonperturbed magnetosphere. Eq. (91) needs a relativistic
correction taking into account the plasma motion: the radiation
density should be multiplied by the relativistic factor $(1 -
v^2/c^2)^{-1/2}$. The net result is
\begin{equation}  \label{92}
W(r,t) = W_0 (r)\Delta^{-11/2}\frac{1}{2}(\Delta^{3/2} +
\Delta^{-3/2}).
\end{equation}
Thus $W(r,t) \sim W_0(r)\Delta^{-7}$. Note a large value of the
exponent of the governing function $\Delta(u)$,which leads to a
large steepness of the local magnetic bremsstrahlung pulse.
Integrating Eq. (92) over the GMSW zone near the magnetic equator,
we obtain a formula for the variation of the complete
magnetospheric magnetic bremsstrahlung resulting from a GW pass:
\begin{equation}  \label{93}
\Delta J(t) = 2\pi
\Theta_0\int\limits_{r_{-}}^{r_{+}}\Phi(\Delta)W_0(r)r^2dr,
\end{equation}
where
\begin{equation}  \label{94}
\Phi(\Delta) = \frac{1}{2}\Delta^{-11/2}(\Delta^{3/2} +
\Delta^{-3/2}) - 1;
\end{equation}
$\Theta_0$
is the angle of the knife radiation pattern. This formula
completely describes the shape of the signal to be detected by a
remote observer. The expression in the square brackets in (93) is
notably nonzero only in the domain of the local GMSW pulse, i.e.
in the domain
\begin{equation}\label{95}
0<S\leq \pi \Leftrightarrow ct-\pi/(\omega\Upsilon(r))\leq r <ct.
\end{equation}

Therefore the integral (93) tends to zero for $t < r_-/c$ and $t >
r_+/c +\pi/(\omega\Upsilon(r))$. The observed pulse duration is
formally determined by these limits. However, since, as noted
above, near the zone boundaries the GMSW is poorly developed, the
actually observed pulse duration (more precisely, its half-width)
can turn out to be smaller than this value. Without solving the
problem of the observed pulse shape, let us estimate its magnitude
in its medium domain
\begin{equation}\label{96}
r_-/c<t<r_+/c,
\end{equation}
when the pulse local duration is much smaller than the window
width:
\begin{equation}\label{97}
\Delta\tau\ll \Delta T.
\end{equation}
Under these conditions the integrand in Eq. (93) is $\delta$-like,
therefore with a good precision the following estimate is valid:
\begin{equation}  \label{98}
\Delta J(t)\approx \frac{2\pi < W_0(r)r^2 > c\Theta_0}{\omega
\Upsilon}\int\limits_{S_{+}}^{S_{-}}\Phi(\Delta(S))dS,
\end{equation}

where $S_{\pm} = \omega \Upsilon(t - r_{\pm}/c), \, < W_0(r)r^2 >$
and the value of $r_*$ is determined from the equation
\begin{equation}\label{99}
r_* = ct- 1/((\omega\Upsilon(r_*)).
\end{equation}
Thus
\begin{equation}\label{100}
\Delta J(t)\sim
\frac{2\pi\Theta_0}{\omega\Upsilon(r_*)}W_0(r_*)r_*^2c\Delta
T\langle\Delta^{-7}\rangle,
\end{equation}
where
\begin{equation}\label{101}
\langle\Delta^{-7}\rangle=\frac{1}{\Delta
S}\int\limits_{S_+}^{S_-}\Delta^{-7}(S)dS;
\end{equation}
$\Delta S=\Delta T\omega\Upsilon$. Thus, by order of magnitude,
the intensity change of the magnetic bremsstrahlung (in its
maximum) as a result of the GMSW excitation is equal to a product
of the unperturbed magnetospheric radiation intensity in the GMSW
zone by the dimensionless factor $\langle\Delta^{-7}\rangle$
(which can reach $10^{-7}$). We will return to a calculation of
the observed pulse shape in our next paper. Here we restrict
ourselves to estimates of the complete energy of a GMSW pulse
(Fig. 16).

\ris{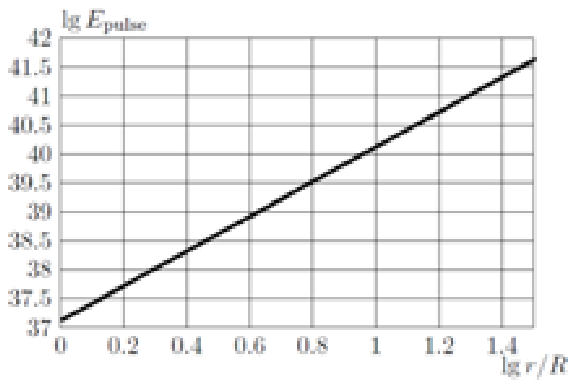}{The radial coordinate ($r$) dependence of the
pulse energy $E_{pulse} = 4\pi r^2 {\cal E}_\Sigma $ (in erg) by
Eq. (80). The GMSW effect is sufficiently well-developed only in
the range $0, 8 < \mbox{lg}(r/R) < 1,2$ ($7,6\cdot 10^6 \mbox{cm}
< r < 1,9 \cdot 10^7 \mbox{cm}$). Outside this region the results
are of clearly formal nature. The pulse energy should rapidly fall
near the boundaries of this range.}

\subsection{Source ofquadrup ole oscillations}
There naturally arises the question of a source ofpulsars’
quadrupole oscillations. A possible energy source for such
oscillations might be represented by explosive nuclear reactions
with heavy hyperons like $n + n \longleftrightarrow p+\Sigma_+$
taking place in neutron stars cores at densities over $10^{15}$
g/cm$^3$ [13]. The presence of strong magnetic fields should lead
to an asymmetry of the explosions, i.e. to the quadrupole moment
excitation. For such processes to occur in a neutron star, it must
be sufficiently young. Numerical simulations of the process of a
neutron star cooling shows [14] that after a Supernova explosion
the neutron star temperature falls approximately by an order of
magnitude in $10^4$ years. Consequently, GMSW should be sought in
radiation from sufficiently young pulsars formed no earlier than
10000 years ago.

\section{The Crab pulsar NP 0532 emits
gravitational waves}
A pulsar with the required parameters does exist: it is the famous
pulsar in the Crab nebula, NP 0532, life-time is less than 1000
years (the 1054 Supernova). This pulsar is the youngest of all
known ones (and consequently the hottest), it has the shortest
period (at least among the closest pulsars, enumerated in Table
2): $T = 0,033$ sec. What is suprising is that the radio emission
of this pulsar contains anomalies which can be with a large degree
of confidence identified with the GMSW. Namely: there are single
irregular, the socalled giant pulses (on the average a pulse in
every 5 to 10 minutes) [11]. The radiation intensity in the giant
pulses is a few tens of times higher (by roughly a factor of 60)
than in common pulses. But the most interesting is that the
duration of the giant pulses is no more than $9\cdot10^{-5}$ s,
i.e., almost by 2(!) orders shorter than that of the common pulses
from NP 0532 ($\tau\sim 6\cdot10^{-3}$ s). The common pulse
duration, as is easily seen, is about 1/7 of the NP 0532 rotation
period, so that the common pulses are clearly explained
geometrically by pulsar rotation. The giant pulse duration is 300
times shorter than the pulsar rotation period, and consequently
the existence of the giant pulses has not yet found any
satisfactory theoretical model.

However, the giant pulses are easily explained by the GMSW, and
their duration is not related to the pulsar rotation period, or an
angle of the knife radiation pattern, but to its eigen-oscillation
period, $T_0$. A comparison of the NP 0532 giant pulse duration
with that of a GMSW pulse (84) shows a striking coincidence.
Indeed, for the NP 0532 pulsar, as known from the annihilation
line shift in the $\gamma$ radiation spectrum (400 keV instead of
511 keV), the gravitational redshift is known \cite{21,22}:
\[
\Delta E/E=MG/Rc^2=0,217.
\]
Then from Table 3 we find the pulsar mass: $M = 1,67M_\oplus$ and
the corresponding neutron star radius: $R$ = 12 km. According to
Eq. (79) and Table 3, the GMSW pulse duration for the NP 0532
pulsar should be about 87 microseconds, while the observed NP 0532
giant pulse duration is approximately 90 microseconds \ref{11}
(!).

For the Crab pulsar $t_0 = 2,5\cdot10^3$ years \ref{15}; then,
setting $\delta M = 1,67$, $R$ = 12 km, we find from (82): $H\sim
5,2\cdot10^{12}$ G. The angle $\Theta_0$ of the knife radiation
pattern is connected with the observed pulse duration $\tau$ by
the relation $\Theta_0=2\pi \tau/P$. For the pulsar NP 0532 this
angle is $0,48\approx 28^o$. Assuming the complete pulsar
luminosity in the continuous spectrum to be about $5\cdot10^{36}$
erg/s, we find the giant pulse intensity recalculated for the
whole neutron star surface:
\[
L_{giant}\approx 4\cdot 10^{39} \mbox{erg/s}.
\]
As has been pointed out above, the real magnetic field strength on
the pulsar NP 0532 surface is $10^{12}$G \cite{11}. To explain the
observed giant pulse emission power, one needs GW magnitude values
on the stellar surface of the order of $10^{-8}$. Note that,
according to Fig. 15, the window width $\Delta T\approx 180 \mu$s
corresponds observed GMSW pulse should be about 90 $\mu$s, which
again precisely coincides with the observed giant pulse duration!

The indicated magnitude $\beta_0(R)$ corresponds to the
gravitational radiation power of the order of $4\cdot 10^{42}$
erg/s and the neutron star oscillations energy about $E_m\approx
4\cdot 10^{41}$ erg. In this case the neutron star surface
oscillation magnitude is about 1 cm. Taking into account that in
the whole lifetime of NP 0532 (1000 years) approximately $7\cdot
10^7$ giant pulses have been emitted, we get an estimate of the
energy carried away from the neutron star by GW for the whole time
of its existence: $E = 2,8 \cdot 10^{49}$ erg. It is $10^{-5}$ of
the rest energy of this neutron star, which completely agrees with
the assumption of a permanent rebuilding of its core. Thus, to a
high degree of confidence we can state that the giant pulses
observed in the pulsar NP 0532 radiation are optical
manifestations of gravimagnetic shock waves (GMSW) excited by the
gravitational radiation of the neutron star corresponding to the
pulsar NP 0532 \cite{20}.

\section{Conclusion}
Besides NP 0532, among all known pulsars only PSR 0833 seems to be
able to emit (but more seldom) giant pulses. Other pulsars are too
old for it. Therefore it is necessary to concentrate the main
effort on observations of these two pulsars. It should be stressed
that there is no other mechanism able to accelerate a shock wave
to subluminal velocities. Therefore an investigation of the giant
pulse spectrum in the X-ray range, aimed at discovering a violet
shift in the radiation spectrum, is of utmost importance. A
comprehensive study of the giant pulses (their shapes and
instantaneous spectrum) will allow one not only to verify the
existence of gravitational radiation, but also to get additional
information on the neutron stars structure and the processes in
their interior. In turn it is necessary to study the GMSW pulse
formation in detail theoretically.

\subsection*{Acknowledgement} The author is thankful to S.V. Sushkov
for help in carring out the calculations and to N.A. Zvereva for
translating the article into English.

\end{document}